\pdfoutput=1
\documentclass[reprint, amsmath, amssymb, aps, nofootinbib]{revtex4-1}
\usepackage[pdftex]{graphicx}
\usepackage{wasysym}
\usepackage{amsfonts}
\usepackage{ulem}
\usepackage{float} %make figure stay

\usepackage[mode=buildnew]{standalone}% requires -shell-escape
\usepackage{enumitem}  

\usepackage[mode=buildnew]{standalone}% requires -shell-escape

\def\2F1{\,_2{\rm F}_1}

\usepackage[usenames,dvipsnames]{xcolor}
\usepackage{color}
\usepackage{graphicx}
\usepackage{dcolumn}
\usepackage{bm}
\usepackage{hyperref}

\newcommand{\beq}{\begin{equation}}
\newcommand{\eeq}{\end{equation}}
\newcommand{\beqq}{\begin{equation*}}
\newcommand{\eeqq}{\end{equation*}}
\newcommand{\bba}{\begin{align}}
\newcommand\beqaa{\begin{eqnarray*}}
\newcommand\eeqaa{\end{eqnarray*}}
\usepackage{tasks}

\usepackage{enumitem} 

\begin{document}

%\null\vskip-10pt \hfill
%\begin{minipage}[t]{42mm}
%SLAC--PUB--16967\\
%\end{minipage}
%\vspace{0mm}

\title{Bootstrapping the \textit{simplest} correlator  in planar $\mathcal{N}=4$ SYM at all loops}

\author{Frank Coronado$^{1,2,3}$}

\affiliation{
\vspace{5mm}
$^{1}$Perimeter Institute for Theoretical Physics, Waterloo, ON N2L 2Y5, Canada\\
$^{2}$Department of Physics $\&$ Astronomy, University of Waterloo, Waterloo, ON N2L 3G1, Canada\\
$^{3}$ICTP South American Institute for Fundamental Research, S\~ao Paulo, SP Brazil 01440-070}

\begin{abstract}

\vspace{3mm}

We present the full form of a four-point correlation function of large BPS operators in planar $\mathcal{N}=4$ Super Yang-Mills to any loop order. We do this by following a bootstrap philosophy based on three simple axioms pertaining to (i) the space of functions arising at each loop order, (ii) the behaviour in the OPE in a double-trace dominated channel and (iii) the behaviour under a double null limit. We discuss how these bootstrap axioms are in turn strongly motivated by empirical observations up to nine loops unveiled through integrability methods in our previous work \cite{MySimplestPaper} on this \textit{simplest} correlation function.

\end{abstract}

\maketitle

\section{Introduction}

\begin{figure}[ht]
\centering
 \resizebox{.5\totalheight}{!}{\includestandalone[width=.8\textwidth]{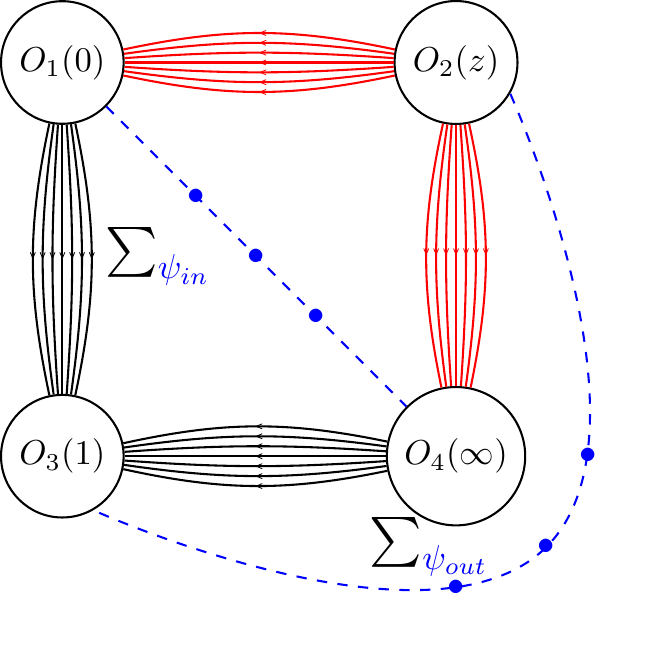}}
 \caption{The \textit{simplest} four-point function with  external operators $O_{1}(0,0) = \text{Tr}(Z^{\frac{K}{2}}\,{\color{red}\bar{X}}^{\frac{K}{2}})+\text{cyclic permutations}$, $O_{2}(z,\bar{z}) = \text{Tr}({\color{red}X}^{K}\,)$, $O_{3}(1,1) = \text{Tr}(\bar{Z}^{K})$ and $O_{4}(\infty,\infty) = \text{Tr}(Z^{\frac{K}{2}}\,{\color{red}\bar{X}}^{\frac{K}{2}}) +\text{cyclic permutations}$. The Wick contractions form a perimeter with four bridges of width $\frac{K}{2}$. According to Hexagonalizaiton \cite{HexagonalizationI} in the limit $K\gg 1$ the loop corrections are obtained by summing over $2D$ intermediate multiparticle states ${\color{blue}\psi_{in}}$ and ${\color{blue}\psi_{out}}$ on mirror cuts $1$-$4$ and $2$-$3$  respectively, with both sums evaluating to $\mathbb{O}$. Alternatively the octagon $\mathbb{O}$ represents the resummation of planar Feynman diagrams draw inside(outside) the perimeter.}
\label{fig:Simplest}
\end{figure}

Integrability methods have shaped a new path for the explicit evaluation of correlators of local operators in planar $\mathcal{N}=4$ SYM \cite{BKV,HexagonalizationI,HexagonalizationII,Cushions,AsymptoticPaper} and also non-planar \cite{NonPlanarI,NonPlanarII,NonPlanarTwoPoint}, specially for  four-point functions of large protected single-trace operators. In \cite{MySimplestPaper} we used integrability-based methods to find the loop corrections to the polarized four-point function we named as the \textit{simplest}. %There we provided explicit results up to nine loops and show how to find arbitrary higher loop corrections.
 This correlator consists of  four external protected operators with $R$-charge polarizations chosen as shown in figure \ref{fig:Simplest}.  In the limit of long operators\footnote{ The rank of the gauge group $N_{c}\to\infty$ is the largest parameter followed by $K$. Then the planar correlator is expanded in powers of the 't  Hooft coupling $g^{2}$.} ($K\gg 1$), we argued this  four-point function admits a factorization into the tree level part which carries all the dependence on the external scaling dimension $K$ and the loop corrections which are given by the squared of the function $\mathbb{O}$ (the octagon)
\beq\label{eq:SimplestFour}
\langle O_{1}O_{2}O_{3}O_{4}\rangle\,=\, \left[\frac{1}{x_{12}^{2}x_{13}^{2}x_{24}^{2}x_{34}^{2}}\right]^{\frac{K}{2}}\times\mathbb{O}^{2}(z,\bar{z}) 
\eeq
where the cross ratios are defined in terms of the spacetime positions as:
\beqq
z\bar{z} = u = \frac{x_{12}^{2}x_{34}^{2}}{x_{13}^{2}x_{24}^{2}}\qquad \text{and} \qquad  (1-z)(1-\bar{z}) = v =\frac{x_{14}^{2}x_{23}^{2}}{x_{13}^{2}x_{24}^{2}}
\eeqq

In this paper we present some of the analytic properties of the octagon $\mathbb{O}$ which follow from the explicit nine-loop results in \cite{MySimplestPaper}.
These properties include a restriction on the space of functions that appear at any loop order and the remarkable simplicity of the octagon in two different kinematical limits: the OPE limit $(z\to 1, \bar{z}\to 1)$  and the double light-cone limit $(z\to 0, \bar{z}\to \infty)$.

We also state that these three analytic properties can be used to uniquely define the octagon and with that also the \textit{simplest} correlator \eqref{eq:SimplestFour}. We show how to solve this bootstrap problem by first introducing a Steinmann basis of Ladders which resolve two of the aforementioned analytic properties. Then using the third property to completely fix the coefficients in an Ansatz constructed with the Steinmann basis.  

This bootstrap approach reproduces the explicit results obtained  from perturbation theory and integrability and  allows us to easily extend them to arbitrary loop order. We accompany this letter with an ancillary file with our explicit results up to 24 loops.

\section{Analytic properties of octagon}
The following analytic properties were observed up to nine loops from the explicit results in \cite{MySimplestPaper}. These empirically found properties will then be converted into bootstrap axioms used to fully determine our correlator. Some of these empirical observations can be a posteriori derived and better understood as discussed in more detail in \cite{TillPedroFrankVascoWorkInProgress}.

\subsection{Single-Valuedness and Ladders}
Our explicit results in \cite{MySimplestPaper} provided the octagon\footnote{here we are refering to the polarized octagon form factor which only depends on spacetime cross ratios $z,\bar{z}$ and has bridge parameter $l=0$. For the more general octagon see \cite{MySimplestPaper}} as a multilinear combination of Ladder integrals:
\beq\label{eq:FinalOctagon}
\mathbb{O}=1+\sum_{n=1}^{\infty}\sum_{J=n^{2}}^{\infty}\sum_{\vec{j}\in Z^{+}_{n}(J)}g^{2J}\times d_{\vec{j}}\times f_{j_{1}}\cdots f_{j_{n}}
\eeq
where $Z^{+}_{n}(J)$ represents the group of sets of positive integers $\vec{j}\equiv \{j_{1},\cdots, j_{n}\}$ which add up to $j_{1}+\cdots +j_{n}=J$. The rational coefficients $d_{\vec{j}}$ are not known in closed form and could be zero for some integer partitions. The basis of conformal Ladder integrals is given by \cite{FirstLadder}
\beqq
f_{p} =-v\sum_{j=p}^{2p}\frac{j!(p-1)!\left[-\log(z\bar{z})\right]^{2p-j}}{(j-p)!(2p-j)!}\left[\frac{\text{Li}_{j}(z)-\text{Li}_{j}(\bar{z})}{z-\bar{z}}\right]
\eeqq
where $v=(1-z)(1-\bar{z})$.

This expansion of $\mathbb{O}$ makes manifest its single-valuedness and its uniform maximal transcendentality at each loop order. 

\subsection{Double-trace OPE channel}
Here we consider the OPE expansion in channel $2$-$3$, see figure \ref{fig:Simplest}. Unlike the other two-channels\footnote{The $1$-$2$ channel was considered in \cite{MySimplestPaper}.  In this channel, at weak coupling, we only get single-traces between the leading twist $K$ and the double-trace threshold $2K$.} ($1$-$2$ and $2$-$4$), this one receives double-trace contributions already at leading twist $2K$. 
 
This OPE limit corresponds to $z\to 1,\bar{z}\to 1$ (or $v\to0,u\to 1$). At weak coupling we find the behaviour of the octagon in  this kinematics to be given by\footnote{Similar truncations have been observed in the study of extremal three-point functions in \cite{PoleBasso}}
\beq\label{eq:Ozto1}
\lim_{z,\bar{z}\to 1}\mathbb{O}(z,\bar{z}) \,=\, \mathsf{a}(z,\bar{z},g^{2}) + \mathsf{b}(z,\bar{z},g^{2}) \,\log v  
\eeq
where both functions $\mathsf{a}$ and $\mathsf{b}$ have a series expansion in the coupling $g^{2}$ and the cross ratios $(1-z)$ and $(1-\bar{z})$. 

In the limit of large operators where the expression \ref{eq:SimplestFour} holds up to arbitrary loop order  this octagon limit \eqref{eq:Ozto1} implies that the \textit{simplest} four-point function has at most a $\log^{2}v$ singularity. This type of truncations is expected in the planar limit for OPE channels dominated by double-trace operators hence we dub this channel as the double trace channel \cite{TillPedroFrankVascoWorkInProgress}.

%.  This type of truncations is expected in the planar limit for OPE channels dominated by double-trace operators\footnote{It is associated to the order on $\frac{1}{N_{c}}$ of the anomalous dimension of this type of operators.} A more detailed study of this OPE channel will be presented in \cite{TillPedroFrankVascoWorkInProgress}.

\subsection{Null-square limit}
This limit corresponds to the kinematics where the external operators become light-like separated: $x_{12}^{2},x_{24}^{2},x_{34}^{2},x_{13}^{2}\to 0$ forming a null square. This limit of the four-point function was considered in \cite{NullCorrelatorWilson} for smaller operators where a relationship where a relationship between null correlators and null polygonal Wilson loops was established.

For our \textit{simplest} four-point function, see \eqref{eq:SimplestFour}, the non-trivial part of this null limit is given by the limit of the octagon\footnote{We also know this limit at strong coupling \cite{TillPedroFrankVascoWorkInProgress} at which the isolated $g^{2}$ term is absent. This makes us believe that its presence in \eqref{eq:LightOctagon} is only a curious artifact of the weak coupling limit.}
\bba\label{eq:LightOctagon}
\lim_{z\to 0,\,\bar{z}\to \infty}\, \log\,\mathbb{O}(z,\bar{z})&=-\tilde{\Gamma}(g)\,\log^{2}(z/\bar{z}) \nonumber\\ &\qquad+\frac{1}{2}\,g^{2}\left(\log^{2}(-z)+\log^{2}(-1/\bar{z})\right)
\end{align}
where the coefficient $\tilde{\Gamma}$  admits an expansion in the coupling
\beqq
\tilde{\Gamma}(g) = \frac{1}{2}\,g^{2} -\frac{1}{6}\,\pi^{2}g^{4} +\frac{8}{45}\,\pi^{4}\,g^{6} - \frac{68}{315}\,\pi^{6}\,g^{8}+\mathcal{O}(g)^{10}
\eeqq
To appreciate better the simplicity of \eqref{eq:LightOctagon} we contrast it against the result for short operators, $K=2$
\begin{itemize}
\item  For the case $K=2$ the coefficient $\tilde{\Gamma}$  is replaced by the cusp anomalous dimension $\Gamma_{cusp}$ which is associated to the energy density of the flux tube between the  Wilson lines. It also appears in the anomalous dimension of the large spin leading twist-operator $tr(ZD^{S}Z)$ dominating the light-cone OPE
\beqq
\Delta  =S\,+\,2 +\,\Gamma_{cusp}(g)\log S +O(1/S)
\eeqq
For our \textit{simplest} correlator the operator(s) dominating the light-cone OPE is of the form $tr(Z^{\frac{K}{2}}D^{S}X^{\frac{K}{2}})$. Furthermore  the limit $K\gg 1$ implies  a huge number of nearly-degenerate operators at leading twist $K$. It would be interesting to analyze how these two latter considerations account for the difference between $\tilde{\Gamma}$ and $\Gamma_{cusp}$. In particular the latter contains odd zeta-numbers while the former only even zeta-numbers.
\item In \eqref{eq:LightOctagon} the exponents of $\log(-z)$ and $\log(-1/\bar{z})$ truncate at degree two while for the case $K=2$ there is an extra complicated function of the cross ratios determined in  \cite{HigherSpinCorrelators} which accounts for the backreaction of the flux-tube on the heavy particle that propagates along the null square, see \cite{NullCorrelatorWilson}.
\end{itemize}
We expect these differences can be explained following an analysis similar to \cite{HigherSpinCorrelators,HigherSpinTowers} including the non-trivial $R$-charge and large $K\gg 1$ limit of our \textit{simplest} correlator \cite{TillPedroFrankVascoWorkInProgress}.  It would also be interesting to see if $\tilde{\Gamma}$ satisfies a linear integral equation as is the case for\footnote{Thanks to B. Basso for comments on this point} $\Gamma_{cusp}$ \cite{LinearIntegralCusp}.

\section{Bootstrapping the octagon}

We now postulate that the analytic properties described in the previous section are valid at all loops and can be used to define a bootstrap problem. More specifically we establish that the perturbative expansion of the \textit{simplest} four-point function is defined by
\begin{enumerate}[label=(\roman*)]
\item  \label{boot:Lad}\textbf{Ladder integrals:} These span the family of functions that appear in the loop corrections of the correlator. They appear in multilinear combinations with uniform maximal transcendentality at any loop order. 
\item \label{boot:Stein} \textbf{Steinmann relations:} The octagon satisfy these relations which establish the vanishing of its double discontinuity
\beq\label{eq:SteinCondition}
\text{Disc}_{1}\text{Disc}_{1}\mathbb{O}(z,\bar{z}) = 0
\eeq
where $\text{Disc}_{1}$ denotes the discontinuity after performing the analytic continuation
$(1-z)\to(1-z)e^{i\pi}$ and $(1-\bar{z})\to(1-\bar{z})e^{i\pi}$. This condition guarantees the truncation to $\log v$ in the OPE expansion $z\to 1,\bar{z}\to1$ at weak coupling.  
\item \label{boot:Light} \textbf{Light-cone asymptotics:} in the null-square limit $z\to 0$ and $z\to \infty$ we demand a simple asymptotics of the logarithm of the octagon:
\bba\label{eq:LightCondition}
\lim_{z\to 0,\bar{z}\to \infty}\;\log\mathbb{O}(z,\bar{z}) \, &= a_{0,0}+a_{1,0}\log(-z)+a_{0,1}\log(-1/\bar{z}) \nonumber\\
&\quad +a_{1,1}\log(-z)\log(-1/\bar{z})\nonumber\\
&\quad+a_{2,0}\log^{2}(-z)+a_{0,2}\log^{2}(-1/\bar{z})
\end{align}
where the relevant condition is the absence of higher $\log$s and we do not impose any conditions on the coefficients $a_{i,j}$.
\end{enumerate} 
In the following sections we show how to resolve these three conditions to determined the octagon and the \textit{simplest} four-point function at any loop order. 
\subsection{A Steinmann basis of Ladder integrals} 
The vanishing of the double discontinuity \ref{boot:Stein} motivates the search for a basis of functions that satisfy this property. Here we combine \ref{boot:Lad} and \ref{boot:Stein} to look for this basis of functions in the space of Ladder integrals. We start with an Ansatz of the form
\beq\label{eq:Sansatz}
\mathcal{S}^{(m,n)}_{i} = \sum_{k_{1}+\cdots + k_{n} =m} d^{(i)}_{k_{1},\cdots , k_{n}}\,f_{k_{1}} \cdots f_{k_{n}}
\eeq
With this Ansatz we are assuming an organization of our Steinmann basis into families $\mathcal{S}^{(m,n)}$  whose elements  have uniform transcendentality of order $m$  and are constructed with $n$ Ladders. We are  provisionally using the sub-index $i$ to label the different elements $\mathcal{S}^{(m,n)}_{i}$ on each family.

In order to find our basis we simply need to take into account the discontinuities of the Ladders: 
\beqaa\label{eq:Double0}
\text{Disc}_{1}\,f^{(n)}(z,\bar{z}) &\sim&   2\pi i
\,\left[\log (z\bar{z})\right]^{n-1} \log\left(\frac{z}{\bar{z}}\right) \\
 \text{Disc}_{1}\text{Disc}_{1}\,f^{(n)}(z,\bar{z}) &=& 0
\eeqaa
then imposing the Steinmann relations
\beq\label{eq:SteinmanCondition}
\text{Disc}_{1}\,\text{Disc}_{1}\,\mathcal{S}^{(m,n)}_{i} = 0
\eeq 
we solve for the coeficients $d$ in the ansatz \eqref{eq:Sansatz}. 

This exercise was performed in \cite{GluingLadders} where some solutions to \eqref{eq:SteinmanCondition} were presented and identified with fishnet  Feynman integrals. Here we will provide all solutions but without a Feynman integral interpretation.
 
We solved equation \eqref{eq:SteinmanCondition} for various $m,n$. From these we gather the following data:
\begin{itemize}
\item  For $m<n^{2}$ there are no solutions. 
\item  For $m=n^{2}$ and $m=n^{2}+1$ there is only one solution.
\item All solutions we found admit determinant representations.
\end{itemize}
This experience allows us to propose a Steinmann basis of Ladders in the form of determinants.  In short, the elements of our Steinmann basis can be identified with the minors of the infinite dimensional matrix
\beqq
\begin{pmatrix}
f_{1} & f_{2}  & f_{3} & \cdots   \\
f_{2}  & f_{3}  & \cdots & \cdots  \\
f_{3}  & \cdots  & \cdots  & \cdots \\
\vdots  & \cdots  & \cdots  & \cdots \\
\end{pmatrix}
\eeqq
more specifically we label these minors as
\beq\label{eq:Mmatrices}
M_{i_{1},i_{2},\cdots, i_{n}} = \begin{vmatrix}
f_{i_{1}} & f_{i_{2}-1}  & \cdots  & f_{i_{n}-n+1} \\
f_{i_{1}+1} & f_{i_{2}} & \cdots  & f_{i_{n}-n+2}\\
\vdots & \vdots & \ddots & \vdots\\
f_{i_{1}+n-1} & f_{i_{2}+n-2} & \dots & f_{i_{n}}\\
\end{vmatrix}
\eeq 
where the subindexes on $M_{i_{1},i_{2},\cdots,i_{n}}$ correspond to the elements on the diagonal and the subindexes on the first row of the matrix  must satisfy
\beqq
0<i_{1}<i_{2}-1<\cdots<i_{n}-n+1
\eeqq
Using these minors we define our Steinmann basis of Ladders as:
\beq\label{eq:SteinmannBasis}
S_{k_{1},\cdots k_{n}} = \left[\prod_{o=1}^{n}\,p_{k_{o}}\right]\,M_{k_{1},\cdots k_{n}}
\eeq
where the rescaling $p_{k}=\frac{1}{k!(k-1)!}$ is just performed for later convenience. The families  $\mathcal{S}^{(m,n)}$ are spanned as follows
\beqq
S_{k_{1},\cdots,k_{n}} \in \mathcal{S}^{(m,n)}\;\;\text{if}\;\;k_{1}+\cdots+k_{n}=m 
\eeqq
Lastly considering the property of maximal transcedentality 
we use our Steinmann basis $\mathcal{S}^{(m,n)}$ to build an Ansatz for each loop order  of a function $\mathbb{O}$ satisfying \ref{boot:Lad} and \ref{boot:Stein}. 
\beq\label{eq:SAnsatz}
\mathbb{O}= 1+\sum_{n=1}^{\infty}\, \sum_{m=n^{2}}^{\infty}\left(g^{2}\right)^{m} \,\sum_{S\in \mathcal{S}^{(m,n)}}\, c_{k_{1},\cdots, k_{n}} \,S_{k_{1},\cdots, k_{n}}
\eeq
\subsection{Fixing all coefficients with Light-cone asymptotics}
In order to fix the coefficients $c_{k_{1},\cdots,k_{n}}$ in the Ansatz we impose the third analytic property \ref{boot:Light}. This condition of exponentiation in the null-square limit allows us to relate coefficients of high loop orders to the ones at lower loops. To take this limit in our Ansatz we simply need to consider  the light-cone limit of the Ladders.
\beq\label{eq:LightLadder}
\lim_{z\to 0,\bar{z}\to \infty}f_{j}(z,\bar{z})=\sum_{m=0}^{j}\sum_{n=0}^{j}\,b^{(j)}_{m,n}\,\log^{m}(-z)\,\log^{n}(-1/\bar{z}) 
\eeq
where $b_{m,n}=0$ if $m+n$ is odd or otherwise:
\beq\label{eq:LightCoe}
b^{(j)}_{m,n} \,=
\,\frac{j!(j-1)!\left(2-2^{m+n-2j+2}\right)\,(2j-m-n)!}{(-1)\,m!\,n!\,(j-m)!\,(j-n)!}\,\zeta_{2j-m-n} \nonumber
\eeq
Notice light-cone Ladder \eqref{eq:LightLadder} is manifestly symmetric under the exchange of cross-ratios $z\leftrightarrow -1/\bar{z}$ and our Ansatz of Ladders directly inherits this feature.

We then enforce the condition of truncation of the exponents of $\log(z)$ and $\log(-1/\bar{z})$ up to degree two. These provides a set of equations which we can be easily solved at each loop order. Up to four-loops the solution looks like:
\beqq\label{eq:4coe}
c_{2} = -2 c_{1}^{2} \,\qquad c_{3} = 6 c_{1}^{3} \,\qquad c_{4} = -20 c_{1}^{4} \qquad c_{1,3} = c_{1}^{4} \nonumber
\eeqq
Likewise we find that we can fix all coefficients $c$ in \eqref{eq:SAnsatz} and $a_{i,j}$ in \eqref{eq:LightCondition} at arbitrary loop order in terms of the single one-loop coefficient $c_{1}$. This latter coefficient can be associated to the definition of the coupling $g^{2}$ and in order to match with the conventions in the literature we set it to $c_{1}=1$. This finally shows that properties \ref{boot:Lad}, \ref{boot:Stein} and \ref{boot:Stein} uniquely define the octagon $\mathbb{O}$ and with that our \textit{simplest} correlator \eqref{eq:SimplestFour} at arbitrary loop order.

Furthermore, we have been able to identify the analytic form of an infinite family of coefficients:
\beqq
c_{\tiny\underbrace{1,3,\cdots, 2n-1}_{n}\,,\, 2n+1+m} = \binom{2m+4 \,n}{m}\qquad \text{with}\quad m\geq 0
\eeqq
In particular the coefficients $c_{1,3,5,\cdots,2n-1}=1$ of the noteworthy elements of our basis $\mathcal{S}_{1,3,5,\cdots,2n-1}$  which have been identified in \cite{GluingLadders} as the fishnet Feynman integrals
\begin{figure}[H]
\centering
 \resizebox{.4\totalheight}{!}{\includestandalone[width=.8\textwidth]{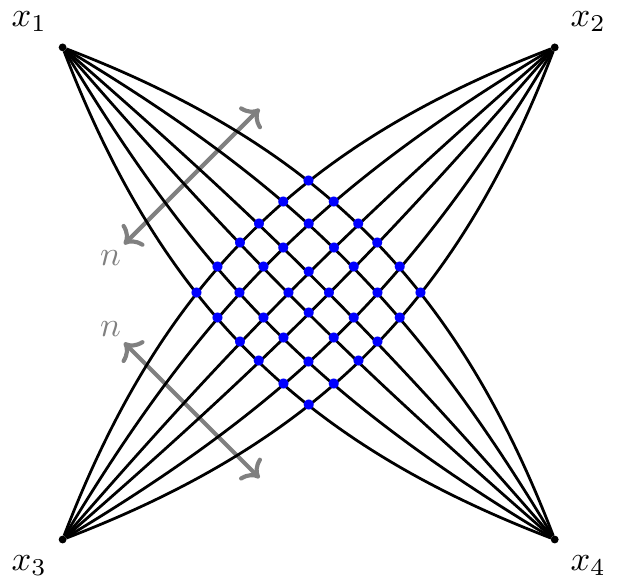}}
\caption{Fishnet identified with $\mathcal{S}_{1,3,\cdots,2n-1}$}
\end{figure}

It is interesting to ask whether other elements of the Steinmann basis of Ladders or perhaps linear combinations of them can be identified with other families of Feynman integrals. Finding such identification could be the guiding principle to find the closed form of all coefficients of our  (possibly rotated) Steinmann basis. Then all would be set to attempt a resummation and get access to the finite or strong coupling limit. This is a question we hope to address in the future.

\section{Conclusion}

In this short letter we have bootstrapped, for the first time in a unitary 4D planar gauge theory, a four-point correlator at all loops in the 't Hooft coupling. This is a correlator of four long protected operators and we call it the \textit{simplest} due to the simplicity of the analytic properties that define it.
These properties, see \ref{boot:Lad} and \ref{boot:Stein}, constrain the space of functions of the loop corrections to a reduced Steinmann basis of Ladders with determinant representations. The coefficients on this basis are then fully determined by imposing a simple exponentiation in the light-cone limit, see \ref{boot:Light}.

An interesting next step is to consider other  kinematical limits of our results. We will be reporting our findings in \cite{TillPedroFrankVascoWorkInProgress}, as well as a more thorough study of the analytic properties presented here and their physical implications.

It would also be interesting to find other higher-point correlation functions that satisfy a version of Steinmann relations. If they exist, finding a basis similar to \eqref{eq:SteinmannBasis} or the Steinmann functions that appear in the context of the S-matrix \cite{Steinmann6,Steinmann7}, would be of relevance to find the loop corrections of these correlators.  A natural candidate would be the six-point correlation function proposed in \cite{MySimplestPaper}, see figure 17 therein.

We also consider important to understant which bootstrap conditions we should include to address the case of generic $R$-charge polarizations and ultimately operators of arbitrary or short scaling dimension. At weak coupling there is a vast list of results, obtained using bootstrap ideas, for the integrands of these correlators \cite{Chicherin:2014uca,Korchemsky:2015ssa,Chicherin:2015bza,3loop,10loop,5loop}. It would be nice to be able to go from the integrand to explicit functions as the ones presented in this paper. 

Recently, bootstrap methods in Mellin space  \cite{Rastelli:2016nze,Rastelli:2017udc} and the analytic conformal bootstrap \cite{Alday:2017xua,Alday:2017vkk,Alday:2018pdi,Caron-Huot:2018kta} have proved fruitful at strong coupling. It would be worthy exploring if these methods can be complemented with bootstrap ideas similar to the ones  presented here to get more results starting in the regime of long operators. 

Finally, it would also be interesting to see if the remarkable analytic properties of the \textit{simplest} correlator also appear in observables of the non-unitary Fishnet theory \cite{FishnetTheory} for which exact correlators have recently been computed \cite{ExactFishnet}.

{\it Acknowledgments:} I thank T.~Bargheer, B.~Basso, T.~Fleury, V.~Goncalves, S.~Komatsu and P.~Vieira for enlightening
discussions and comments on the manuscript.

\end{document}